\documentclass[11pt]{article}
\usepackage{moriond,epsfig}

\usepackage{amsfonts,amsmath,amssymb,bm}

\def\be{\begin{equation}}
\def\ee{\end{equation}}
\def\bea{\begin{eqnarray}}
\def\eea{\end{eqnarray}}

\let\th\theta
\newcommand{\vecnab}{\bm{\nabla}}
\newcommand{\ud}{\mathrm{d}}

\begin{document}
\vspace*{4cm}
\title{Testing MOND in the Solar System}

\author{Luc BLANCHET}

\address{$\mathcal{G}\mathbb{R}\varepsilon{\mathbb{C}}\mathcal{O}$, Institut d'Astrophysique de Paris, CNRS, \\Universit\'e Pierre
  et Marie Curie, 98$^{\rm bis}$ boulevard Arago, 75014 Paris, France}

\author{J\'er\^ome NOVAK}

\address{Laboratoire Univers et Th\'eories, Observatoire de
  Paris, CNRS, \\Universit\'e Denis Diderot, 5 place Jules Janssen,
  92190 Meudon, France}

\maketitle\abstracts{The Modified Newtonian Dynamics (MOND) generically predicts a violation of the strong version of the equivalence principle. As a result the gravitational dynamics of a system depends on the external gravitational field in which the system is embedded. This so-called external field effect is shown to imply the existence of an anomalous quadrupolar correction, along the direction of the external galactic field, in the gravitational potential felt by planets in the Solar System. We compute this effect by a numerical integration of the MOND equation in the presence of an external field, and deduce the secular precession of the perihelion of planets induced by this effect. We find that the precession effect is rather large for outer gaseous planets, and in the case of Saturn is comparable to, and in some cases marginally excluded by published residuals of precession permitted by the best planetary ephemerides.}

\section{The external field effect with MOND} 
\label{s:intro}

The Modified Newtonian Dynamics (MOND) has been proposed~\cite{Milg} as an alternative to the dark matter paradigm~\cite{SandMcG02}. At the non-relativistic level, the best formulation of MOND is the modified Poisson equation~\cite{BekM84},
\begin{equation}\label{e:MOND}
\vecnab \cdot \left[ \mu\left(\frac{g}{a_0}\right) \vecnab U
  \right] = -4\pi G \rho\,,  
\end{equation}
where $\rho$ is the density of ordinary (baryonic) matter, $U$ is the gravitational potential, $\bm{g}=\vecnab U$ is the gravitational field and $g = \vert\bm{g}\vert$ its ordinary Euclidean norm. The modification of the Poisson equation is encoded in the MOND function $\mu(y)$ of the single argument $y\equiv g/a_0$, where $a_0=1.2\times 10^{-10}\,\mathrm{m}/\mathrm{s}^2$ denotes the MOND constant acceleration scale. The MOND function interpolates between the MOND regime corresponding to weak gravitational fields $y=g/a_0\ll 1$, for which it behaves as $\mu(y)=y+o(y)$, and the Newtonian strong-field regime $y\gg 1$, where $\mu$ reduces to $1$ so that we recover the usual Newtonian gravity.

An important consequence of the non-linearity of Eq.~\eqref{e:MOND} in the MOND regime, is that the gravitational dynamics of a system is influenced (besides the well-known tidal force) by the external gravitational environment in which the system is embedded. This is known as the external field effect (EFE), which has non-trivial implications for non-isolated gravitating systems. The EFE was conjectured to explain the dynamics of open star clusters in our galaxy~\cite{Milg}, since they do not show evidence of dark matter despite the involved weak internal accelerations (i.e. below $a_0$). The EFE effect shows that the dynamics of these systems should actually be Newtonian as a result of their immersion in the gravitational field of the Milky Way. The EFE is a rigorous prediction of the equation \eqref{e:MOND}, and is best exemplified by the asymptotic behaviour of the solution of \eqref{e:MOND} far from a localised matter distribution (say, the Solar System), in the presence of a constant external gravitational field $\bm{g}_\text{e}$ (the field of the Milky Way). At large distances $r=\vert\mathbf{x}\vert\to\infty$ we have~\cite{BekM84}
\begin{equation}
  \label{e:asym}
  U = \bm{g}_\text{e}\cdot\mathbf{x} + \frac{GM/\mu_\text{e}}{r \sqrt{1 + \lambda_\text{e} \sin^2 \th}} + \mathcal{O} \left(
    \frac{1}{r^2} \right)\,,
\end{equation}
where $M$ is the mass of the localised matter distribution, where $\th$ is the polar angle from the direction of the external field $\bm{g}_\text{e}$, and where we denote $\mu_\text{e} \equiv \mu(y_\text{e})$ and $\lambda_\text{e} \equiv y_\text{e} \mu'_\text{e}/\mu_\text{e}$, with $y_\text{e}=g_\text{e}/a_0$ and $\mu'_\text{e}= \ud\mu(y_\text{e})/\ud y_\text{e}$. In the presence of the external field, the MOND internal potential $u\equiv U-\bm{g}_\text{e}\cdot\mathbf{x}$ shows a Newtonian-like fall-off $\sim r^{-1}$ at large distances but with an effective gravitational constant $G/\mu_\text{e}$.\,\footnote{Recall that in the absence of the external field the MOND potential behaves like $U\sim -\sqrt{GM a_0}\ln r$, showing that there is no escape velocity from an isolated system~\cite{FBZ08}. However since no object is truly isolated the asymptotic behaviour of the potential is always given by \eqref{e:asym}, in the approximation where the external field is constant.} However, contrary to the Newtonian case, it exhibits a non-spherical deformation along the direction of the external field. The fact that the external field $\bm{g}_\text{e}$ does not disappear from the internal dynamics can be interpreted as a violation of the strong version of the equivalence principle.

\section{Abnormal influence of the Galaxy in the Solar System}

In two recent papers~\cite{Milg09,BN11} it was shown that the imprint of the external galactic field $\bm{g}_\text{e}$ on the Solar System (due to a violation of the strong equivalence principle) shows up not only asymptotically, but also in the inner regions of the system, where it may have implications for the motion of planets. This is somewhat unexpected because gravity is strong there (we have $g\gg a_0$) and the dynamics should be Newtonian. However, because of the properties of the equation \eqref{e:MOND}, the solution will be given by some non-local Poisson integral, and the dynamics in the strong-field region will be affected by the anomalous behaviour in the asymptotic weak-field region. 

We assume that the external Galactic field $\bm{g}_\text{e}$ is constant over the entire Solar System.\,\footnote{For the Milky Way field at the level of the Sun we have $g_\text{e}\simeq 1.9\times 10^{-10}\,\mathrm{m}/\mathrm{s}^2$ which happens to be slightly above the MOND scale, i.e. $\eta\equiv g_\text{e}/a_0  \simeq 1.6$.} The motion of planets of the Solar System relatively to the Sun obeys the internal gravitational potential $u$ defined by
\begin{equation}
  \label{e:def_u}
  u = U - \bm{g}_\text{e} \cdot \mathbf{x}\,,
\end{equation}
which is such that $\lim_{r\to \infty}u = 0$. Contrary to what happens in the Newtonian case, the external field $\bm{g}_\text{e}$ does not disappear from the gravitational field equation \eqref{e:MOND} and we want to investigate numerically its effect. The anomaly detected by a Newtonian physicist is the difference of internal potentials,
\begin{equation}
  \label{uuN}
\delta u = u - u_\text{N}\,,
\end{equation}
where $u_\mathrm{N}$ denotes the ordinary Newtonian potential generated by the same ordinary matter distribution $\rho$, and thus solution of the Poisson equation $\Delta u_\text{N}= -4\pi G \rho$ with the boundary condition that $\lim_{r\to \infty}u_\text{N} = 0$. We neglect here the change in the matter distribution $\rho$ when considering MOND theory instead of Newton's law. This is in general a good approximation because the gravitational field giving the hydrostatic equilibrium (and thus $\rho$) is strong and MOND effects are very small. Hence $u_\mathrm{N}$ is given by the standard Poisson integral. 

A short calculation shows that the anomaly obeys the Poisson equation $\Delta \delta u = -4\pi G \rho_\text{pdm}$, where $\rho_\text{pdm}$ is the density of ``phantom dark matter'' defined by
\begin{equation}
  \label{pdm}
\rho_\text{pdm} = \frac{1}{4\pi G}\vecnab\cdot\left(\chi\vecnab U\right)\,,
\end{equation}
where we denote $\chi\equiv\mu-1$. The phantom dark matter represents the mass density that a Newtonian physicist would attribute to dark matter. In the model~\cite{BL08,BL09} the phantom dark matter is interpreted as the density of polarisation of some dipolar dark matter medium and the coefficient $\chi$ represents the ``gravitational susceptibility'' of this dark matter medium.

The Poisson equation $\Delta \delta u = -4\pi G \rho_\text{pdm}$ is to be solved with the boundary condition that $\lim_{r\to \infty}\delta u = 0$; hence the solution is given by the Poisson integral
\begin{equation}
  \label{deltaupoisson}
\delta u(\mathbf{x},t) = G \int \frac{\ud^3 \mathbf{x}'}{\vert\mathbf{x}-\mathbf{x}'\vert}\,\rho_\text{pdm}(\mathbf{x}',t)\,.
\end{equation}
We emphasise that, contrary to the Newtonian (linear) case, the knowledge of the matter density distribution does not allow to obtain an analytic solution for the potential, and the solution has to be investigated numerically. We can check that the phantom dark matter behaves like $r^{-3}$ when $r\to \infty$, so the integral \eqref{deltaupoisson} is perfectly convergent.

In the inner part of the Solar System the gravitational field is strong ($g\gg a_0$) thus $\mu$ tends to one there, and $\chi$ tends to zero. Here we adopt the extreme case where $\chi$ is \textit{exactly} zero in a neighbourhood of the origin, say for $r\leqslant\varepsilon$, so that there is no phantom dark matter for $r\leqslant\varepsilon$; for the full numerical integration later we shall still make this assumption by posing $\chi=0$ inside the Sun (in particular we shall always neglect the small MOND effect at the centre of the Sun where gravity is vanishingly small). If $\rho_\text{pdm}=0$ when $r\leqslant\varepsilon$ we can directly obtain the multipolar expansion of the anomalous term \eqref{deltaupoisson} about the origin by Taylor expanding the integrand when $r=\vert\mathbf{x}\vert\to 0$. In this way we obtain\,\footnote{Our notation is as follows: $L = i_1 \cdots i_l$ denotes a multi-index composed of $l$ multipolar spatial indices $i_1, \cdots, i_l$ (ranging from 1 to 3); $\partial_L = \partial_{i_1} \cdots \partial_{i_l}$ is the product of $l$ partial derivatives $\partial_i \equiv \partial / \partial x^i$; $x^L = x^{i_1} \cdots x^{i_l}$ is the product of $l$ spatial positions $x^i$; similarly $n^L = n^{i_1} \cdots n^{i_l} = x^L/r^l$ is the product of $l$ unit vectors $n^i=x^i/r$; the symmetric-trace-free (STF) projection is indicated with a hat, for instance $\hat{x}^L \equiv \text{STF}[x^L]$, and similarly for $\hat{n}^L$ and $\hat{\partial}_L$. In the case of summed-up (dummy) multi-indices $L$, we do not write the $l$ summations from 1 to 3 over their indices.} 
\begin{equation}
  \label{multexp}
\delta u = \sum_{l=0}^{+\infty} \frac{(-)^l}{l!} \,x^L Q_L\,,
\end{equation}
where the multipole moments near the origin are given by
\begin{equation}
  \label{QLpdm}
Q_L = G \int_{r > \varepsilon} \ud^3 \mathbf{x} \,\rho_\text{pdm}\,\partial_L\left(\frac{1}{r}\right)\,.
\end{equation}
Because the integration in \eqref{QLpdm} is limited to the domain $r>\varepsilon$ and $\partial_L(1/r)$ is symmetric-trace-free (STF) there [indeed $\Delta(1/r)=0$], we deduce that the multipole moments $Q_L$ themselves are STF. This can also be immediately inferred from the fact that $\Delta \delta u = 0$ when $r\leqslant\varepsilon$, hence the multipole expansion \eqref{multexp} must be a homogeneous solution of the Laplace equation which is regular at the origin, and is therefore necessarily made solely of STF tensors of type $\hat{x}^L$. Hence we can replace $x^L$ in \eqref{multexp} by its STF projection $\hat{x}^L$. It is now clear from the non-local integral in \eqref{QLpdm} that the MONDian gravitational field (for $r \geqslant r_0$) can influence the near-zone expansion of the field when $r \to 0$. An alternative expression of the multipole moments can also be proved, either directly or by explicit transformation of the integral \eqref{QLpdm}. We have
\begin{equation}
  \label{QLres}
Q_L = - u_\text{N}(\mathbf{0})\,\delta_{l,0} + (-)^l (\hat{\partial}_L u)(\mathbf{0})\,,
\end{equation}
where the Newtonian potential $u_\text{N}$ and the STF derivatives of the internal potential $u$ are to be evaluated at the centre $\mathbf{0}$ of the Sun. 

The multipole expansion \eqref{multexp} will be valid whenever $r$ is much less than the MOND transition distance for the Solar System, defined by $r_0=\sqrt{G M/a_0}$ with $M$ the mass of the Sun and $a_0$ the MOND acceleration scale. This radius corresponds to the transition region where the Newtonian acceleration becomes of the order of the MOND acceleration $a_0$ and therefore, MOND effects become dominant. We have $r_0\simeq 7100\,\mathrm{AU}$ so the results \eqref{multexp}--\eqref{QLres} hold in a large volume around the Sun including all the planets (recall that Neptune's orbit is at $30\,\mathrm{AU}$).

\section{Results for the induced quadrupole moment in the Solar System} 

So far we have elucidated the structure of the multipole expansion of the anomaly $\delta u$ near the origin. Next we resort to a numerical integration of the non-linear MOND equation \eqref{e:MOND} in order to obtain quantitative values for the multipole moments.\,\footnote{Our numerical scheme is based on the very efficient integrator of elliptic equations \textsc{lorene}, available from the website \texttt{http://www.lorene.obspm.fr}.}
\begin{figure*}
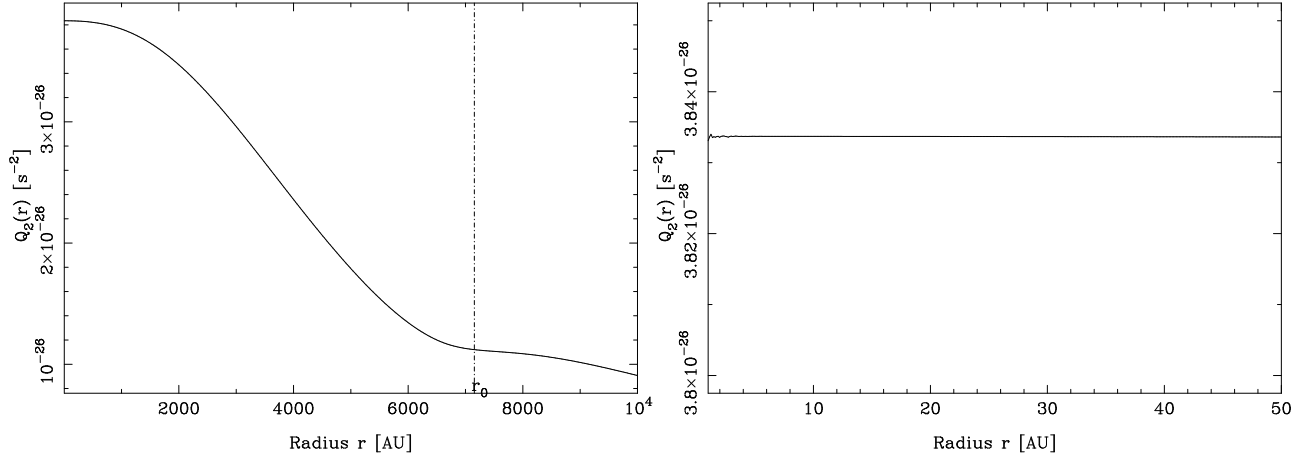

  \centerline{\includegraphics[width=6cm,angle=-90]{q_r_alpha1.ps}\includegraphics[width=6cm,angle=-90]{qr_zoom.ps}}
  \caption{Left panel: profile of $Q_2(r)$ in the Solar System, for a standard choice of function $\mu_1(y)$ [see Eq.~\eqref{mun}], $a_0 = 1.2 \times 10^{-10} \textrm{m.s}^{-2}$ and $g_\text{e} = 1.9 \times 10^{-10} \textrm{ m.s}^{-2}$. The MOND transition radius is shown by a dash-dotted line at $r_0\simeq 7100$ AU. Right panel: zoom of the central region ($r\leqslant 50$ AU), where the quadrupole is almost constant.}
  \label{f:qralpha1}
\end{figure*}

The Sun being assumed to be spherically symmetric, since all the multipole moments are induced by the presence of the external field $\bm{g}_\text{e}$ in the preferred direction $\bm{e}$, the situation is axisymmetric and all the moments $Q_L$ will have their axis pointing in that direction $\bm{e}$. Thus we can define some multipole coefficients $Q_l$ by posing $Q_L = Q_l \,\hat{e}^L$, where $\hat{e}^L$ denotes the STF part of the product of $l$ unit vectors $e^L=e^{i_1}\cdots e^{i_l}$. The multipole expansion \eqref{multexp} reads then as
\begin{equation}
  \label{legendre}
\delta u(r, \th) = \sum_{l=0}^{+\infty} \frac{(-)^l}{(2l-1)!!} \,r^l \,Q_l(r) \,P_l (\cos\th)\,,
\end{equation}
where $P_l(z)$ is the usual Legendre polynomial and $\th$ is the angle away from the Galactic direction $\bm{e}$. Although from the previous considerations the multipole coefficients $Q_l$ should be approximately constant within the MOND transition radius $r_0$, here we compute them directly from the numerical solution of \eqref{e:MOND} and shall obtain their dependence on $r$. With our definition the quadrupolar piece in the internal field is given by
\begin{equation}\label{e:def_quadrupole}
\delta u_2 = \frac{1}{2} \,r^2 \,Q_2(r) \left(\cos^2\th - \frac{1}{3}\right)\,.
\end{equation}
The radial dependence of the anomaly \eqref{e:def_quadrupole} is $\propto r^2$ and can thus be separated from a quadrupolar deformation due to the Sun's oblateness which decreases like $\propto r^{-3}$.

As a first result, we show in Fig.~\ref{f:qralpha1} the profile of the quadrupole induced by the MOND theory through the function $Q_2(r)$ defined in Eq.~\eqref{e:def_quadrupole}. We find that this quadrupole is decreasing from the Sun's neighbourhood to zero, on a typical scale of $10000$ astronomical units (AU). However, we check numerically that $Q_2(r)$ is almost constant in a large sphere surrounding the Solar system, as it has a relative variation lower than $10^{-4}$ within 30 AU (see the zoomed region in Fig.~\ref{f:qralpha1}). We shall therefore refer to the quadrupole as a simple number, noted $Q_2(0)$ or simply $Q_2$, when evaluating its influence on the orbits of Solar-system planets. 

On dimensional analysis we expect that the quadrupole coefficient $Q_2$ should scale with the MOND acceleration $a_0$ like
\begin{equation}\label{q2}
Q_2 = \frac{a_0}{r_0} \,q_2(\eta)\,,
\end{equation}
where $r_0=\sqrt{G M/a_0}$ is the MOND transition radius and where the dimensionless coefficient $q_2$ depends on the ratio $\eta=g_\text{e}/a_0$ between the external field and $a_0$, and on the choice of the interpolating function $\mu$. Our numerical results for the quadrupole are given in Table~\ref{tab1}, for different coupling functions $\mu(y)$.\,\footnote{Note that the quadrupole coefficient $Q_2$ is found to be always positive which corresponds to a prolate elongation along the quadrupolar axis.} Here we consider various cases widely used in the literature:
\begin{subequations}\label{mu}
\begin{align}
\mu_n(y) &= \frac{y}{\sqrt[n]{1+y^n}}\,,\label{mun}\\
\mu_{\textrm{exp}}(y) &= 1-e^{-y}\,,\label{mu3}\\
\mu_{\textrm{TeVeS}}(y) &= \dfrac{\sqrt{1+4y} - 1}{\sqrt{1+4y} + 1}\,.\label{mu4}
\end{align}
\end{subequations}
The function $\mu_1$ has been shown to yield good fits of galactic rotation curves~\cite{FB05}; However because of its slow transition to the Newtonian regime it is \textit{a priori} incompatible with Solar System observations. The function $\mu_2$ is generally called the ``standard'' choice and was used in fits~\cite{BBS91}. We include also the function $\mu_{\textrm{exp}}$ having an exponentially fast transition to the Newtonian regime. The fourth choice $\mu_{\textrm{TeVeS}}$ is motivated by the TeVeS theory~\cite{Bek04}. One should note that none of these functions derives from a fundamental physical principle.
\begin{table*}
\centering
  \caption{Numerical values of the quadrupole $Q_2$ together with the associated dimensionless quantity $q_2$ defined by Eq.~\eqref{q2}. All values are given near the Sun. We use different choices of the function $\mu(y)$ defined in Eqs.~\eqref{mu}.}
\vspace{0.2cm}
    \label{tab1}
  \begin{tabular}{lccccc}
\hline
MOND function & $\mu_1(y)$ & $\mu_2(y)$ & $\mu_{20}(y)$ & $\mu_{\textrm{exp}}(y)$ & 
    $\mu_{\textrm{TeVeS}}(y)$\\
    \hline
    $Q_2$ [$\text{s}^{-2}$] & $3.8\times 10^{-26}$ & $2.2\times 10^{-26}$ & $2.1\times10^{-27}$& $3.0\times 10^{-26}$ &$4.1\times 10^{-26}$ \\
    $q_2$ & $0.33$ &$0.19$ & $1.8\times 10^{-2}$ & $0.26$ &$0.36$\\
\hline
  \end{tabular}
\end{table*}

We have used several functions of type $\mu_n$, as defined in Eq.~\eqref{mun}. One can notice that the value of $Q_2$ decreases with $n$, that is with a faster transition from the weak-field regime where $\mu(y) \sim y$, to the strong field regime where $\mu(y) \sim 1$. We have been unable to determine numerically a possible limit for $Q_2$ as $n$ goes to infinity.

\section{Effect on the dynamics of the Solar System planets}\label{s:effect}

We investigate the consequence for the dynamics of inner planets of the Solar System of the presence of an abnormal quadrupole moment $Q_2$ oriented toward the direction $\bm{e}$ of the galactic centre. Recall that the domain of validity of this anomaly is expected to enclose all the inner Solar System (for distances $r\lesssim r_0\approx 7100$ AU), with the quadrupole coefficient being constant up to say $50$ AU (see Fig.~\ref{f:qralpha1}). As we have seen, the anomaly induces a perturbation on the Newtonian gravitational potential, namely $u=u_\text{N}+\delta u$, where $u_\text{N}=G M/r$ and the perturbation function $R\equiv\delta u$ is given for the quadrupole moment by Eq.~\eqref{e:def_quadrupole}.

We apply the standard linear perturbation equations of celestial mechanics~\cite{BrouwerClemence}. The unperturbed Keplerian orbit of a planet around the Sun is described by six orbital elements. For these we adopt the semi-major axis $a$, the eccentricity $e$, the inclination $I$ of the orbital plane, the mean anomaly $\ell$ defined by $\ell=n(t-T)$ where $n=2\pi/P$ ($n$ is the mean motion, $P$ is the orbital period and $T$ is the instant of passage at the perihelion), the argument of the perihelion $\omega$ (or angular distance from ascending node to perihelion), and the longitude of the ascending node $\Omega$. We also use the longitude of the perihelion defined by $\tilde{\omega}=\omega+\Omega$.

The perturbation function $R=\delta u_2$ is a function of the orbital elements of the unperturbed Keplerian ellipse, say $\{c_A\}=\{a,e,I,\ell,\omega,\Omega\}$. The perturbation equations are generated by the partial derivatives of the perturbation function with respect to the orbital elements, namely $\partial R/\partial c_A$. We express the planet's absolute coordinates $(x,y,z)$ (in some absolute Galilean frame) in terms of the orbital elements $\{a,e,I,\ell,\omega,\Omega\}$ by performing as usual three successive frame rotations with angles $\Omega$, $I$ and $\omega$, to arrive at the frame $(u,v,w)$ associated with the motion, where $(u,v)$ is in the orbital plane, with $u$ in the direction of the perihelion and $v$ oriented in the sense of motion at perihelion. The unperturbed coordinates of the planet in this frame are
\begin{subequations}\label{comp}
\begin{eqnarray}
u &=& a \left(\cos U -e\right)\,,\\
v &=& a \sqrt{1-e^2} \sin U\,,\\
w &=& 0\,,
\end{eqnarray}\end{subequations}
where $U$ denotes the eccentric anomaly, related to $\ell$ by the Kepler equation $\ell = U - e \sin U$. The perturbation equations provide the variations of the orbital elements $\ud c_A/\ud t$ as linear combinations of the partial derivatives $\partial R/\partial c_B$ of the perturbation function. We are interested only in secular effects, so we average in time the perturbation equations over one orbital period $P$. Denoting the time average by brackets, and transforming it to an average over the eccentric anomaly $U$, we have
\begin{equation}\label{av}
\left\langle\frac{\ud c_A}{\ud t}\right\rangle = \frac{1}{P} \int_0^P \ud t \,\frac{\ud c_A}{\ud t} = \frac{1}{2\pi}\int_0^{2\pi} \ud U \,\left(1 - e \cos U\right)\,\frac{\ud c_A}{\ud t}\,.
\end{equation}

In the following, to simplify the presentation, we shall choose the $x$-direction of the absolute Galilean frame to be the direction of the galactic centre $\bm{e}=\bm{g}_\text{e}/g_\text{e}$. That is, we assume that the origin of the longitude of the ascending node $\Omega$ lies in the direction of the galactic centre. Furthermore, in order to make some estimate of the magnitude of the quadrupole effect, let us approximate the direction of the galactic centre (which is only $5.5$ degrees off the plane of the ecliptic) as being located in the plane of the orbit; consequently we choose $I=0$. In this case $\tilde{\omega}=\omega+\Omega$ is the relevant angle for the argument of the perihelion. We then find the following non-zero evolution equations: 
\begin{subequations}\label{avperteqsI0}
\begin{eqnarray}
\left\langle\frac{\ud e}{\ud t}\right\rangle &=& \frac{5 Q_2 e \sqrt{1-e^2}}{4n} \sin (2\tilde{\omega})\,,\\
\left\langle\frac{\ud \ell}{\ud t}\right\rangle &=& n - \frac{Q_2}{12 n}\Bigl[ 7 + 3 e^2 + 15 (1+e^2) \cos (2\tilde{\omega})\Bigr]\,,\\
\left\langle\frac{\ud \tilde{\omega}}{\ud t}\right\rangle &=& \frac{Q_2 \sqrt{1-e^2}}{4 n}\Bigl[ 1 + 5 \cos (2\tilde{\omega})\Bigr]\,.
\end{eqnarray}\end{subequations}
We recall that $\tilde{\omega}$ is the azimuthal angle between the direction of the perihelion and that of the galactic centre (approximated to lie in the orbital plane). Of particular interest is the secular precession of the perihelion $\langle\ud \tilde{\omega}/\ud t\rangle$ due to the quadrupole effect henceforth denoted by
\begin{equation}\label{Delta2}
\Delta_2 = \frac{Q_2 \sqrt{1-e^2}}{4 n}\Bigl[ 1 + 5 \cos (2\tilde{\omega})\Bigr]\,.
\end{equation}
The precession is non-spherical, in the sense that it depends on the orientation of the orbit relative to the galactic centre through its dependence upon the perihelion's longitude $\tilde{\omega}$. The effect scales with the inverse of the orbital frequency $n=2\pi/P$ and therefore becomes more important for outer planets like Saturn than for inner planets like Mercury. This is in agreement with the fact that the quadrupole effect we are considering increases with the distance to the Sun (but of course will fall down when $r$ becomes appreciably comparable to $r_0$, see Fig.~\ref{f:qralpha1}). 

\begin{table*}
\small
\begin{center}
\caption{Results for the precession rates of planets $\Delta_2$ due to the quadrupole coefficient $Q_2$. We use the values for $Q_2$ for various MOND functions as computed in Table~\ref{tab1}. Published postfit residuals of orbital precession (after taking into account the relativistic precession). All results are given in milli-arc-seconds per century.}
\label{tab2}
\begin{tabular}{ccccccccc}
\hline
\multicolumn{9}{|c|}{Quadrupolar precession rate $\Delta_2$ in $\text{mas}/\text{cy}$}\\
\hline
\hspace{-0.3cm}MOND function & \hspace{-0.3cm}Mercury & \hspace{-0.3cm}Venus & \hspace{-0.3cm}Earth & \hspace{-0.3cm}Mars & \hspace{-0.3cm}Jupiter & \hspace{-0.3cm}Saturn & \hspace{-0.3cm}Uranus & \hspace{-0.3cm}Neptune\\
\cline{2-9}
\hline
	$\mu_1(y)$ & $0.04$ & $0.02$ & $0.16$ & $-0.16$ & $-1.12$ & $5.39$ & $-10.14$ & $7.93$ \\
	$\mu_2(y)$ & $0.02$ & $0.01$ & $0.09$ & $-0.09$ & $-0.65$ & $3.12$ & $-5.87$ & $4.59$ \\
	$\mu_{20}(y)$ & $2\times 10^{-3}$ & $10^{-3}$ & $9\times 10^{-3}$ & $-9\times 10^{-3}$ & $-0.06$ & $0.3$ & $-0.56$ & $0.44$ \\
	$\mu_{\textrm{exp}}(y)$ & $0.03$ & $0.02$ & $0.13$ & $-0.13$ & $-0.88$ & $4.25$ & $-8.01$ & $6.26$ \\
	$\mu_{\textrm{TeVeS}}(y)$ & $0.05$ & $0.02$ & $0.17$ & $-0.17$ & $-1.21$ & $5.81$ & $-10.94$ & $8.56$ \\
\hline
\multicolumn{9}{|c|}{Postfit residuals for $\Delta=\langle\ud\tilde{\omega}/\ud t\rangle$ in $\text{mas}/\text{cy}$}\\
\hline
	\hspace{-0.2cm}Origin & \hspace{-0.2cm}Mercury & \hspace{-0.2cm}Venus & \hspace{-0.2cm}Earth & \hspace{-0.2cm}Mars & \hspace{-0.3cm}Jupiter & \hspace{-0.2cm}Saturn & \hspace{-0.2cm}Uranus & \hspace{-0.2cm}Neptune\\
\hline
	Pitjeva~\cite{pitjeva05} & \hspace{-0.2cm}$-3.6\pm 5$ & \hspace{-0.2cm}$-0.4\pm 0.5$ & \hspace{-0.2cm}$-0.2\pm 0.4$ & \hspace{-0.2cm}$0.1\pm 0.5$ & - & \hspace{-0.2cm}$-6\pm 2$ & - & - \\
	Fienga \textit{et al.}~\cite{fienga09} & \hspace{-0.2cm}$-10\pm 30$ & \hspace{-0.2cm}$-4\pm 6$ & \hspace{-0.2cm}$0\pm 0.016$ & \hspace{-0.2cm}$0\pm 0.2$ & \hspace{-0.3cm}$142\pm 156$ & \hspace{-0.2cm}$-10\pm 8$ & \hspace{-0.2cm}$0\pm 2\cdot10^4$ & \hspace{-0.2cm}$0\pm 2\cdot10^4$ \\
	Fienga \textit{et al.}~\cite{fienga10} & \hspace{-0.2cm}$0.4\pm 0.6$ & \hspace{-0.2cm}$0.2\pm 1.5$ & \hspace{-0.2cm}$-0.2\pm 0.9$ & \hspace{-0.2cm}$-0.04\pm 0.15$ & \hspace{-0.3cm}$-41\pm 42$ & \hspace{-0.2cm}$0.15\pm 0.65$ & \hspace{-0.2cm} - & \hspace{-0.2cm} -\\
\hline
\end{tabular}
\end{center}
\end{table*}
Our numerical values for the quadrupole anomalous precession $\Delta_2$ are reported in Table~\ref{tab2}. As we see the quadrupolar precession $\Delta_2$ is in the range of the milli-arc-second per century which is not negligible. In particular it becomes interestingly large for the outer gaseous planets of the Solar System, essentially Saturn, Uranus and Neptune. The dependence on the choice of the MOND function $\mu$ is noticeable only for functions $\mu_n(y)$ defined by \eqref{mun} with large values of $n$, where the effect decreases by a factor $\sim 10$ between $n=2$ and $n=20$.

We then compare in Table~\ref{tab2} our results to the best published postfit residuals for any possible supplementary precession of planetary orbits (after the relativistic precession has been duly taken into account), which have been obtained from global fits of the Solar System dynamics~\cite{pitjeva05,fienga09,fienga10}. In particular the postfit residuals obtained by the INPOP planetary ephemerides~\cite{fienga09,fienga10} use information from the combination of very accurate tracking data of spacecrafts orbiting different planets. We find that the values for $\Delta_2$ are smaller or much smaller than the published residuals except for the planets Mars and Saturn. Very interestingly, our values are smaller or grossly within the range of the postfit residuals for these planets. In the case of Saturn notably, the constraints seem already to exclude most of our obtained values for $\Delta_2$, except for MOND functions of the type $\mu_n$ and given by \eqref{mun} with rather large values of $n$. 

However let us note that the INPOP ephemerides are used to detect the presence of an eventual abnormal precession, not to adjust precisely the value of that precession~\cite{fienga09,fienga10}. On the other hand the postfit residuals are obtained by adding by hands an excess of precession for the planets and looking for the tolerance of the data on this excess~\cite{fienga09,fienga10}. But in order to really test the anomalous quadrupolar precession rate $\Delta_2$, one should consistently work in a MOND picture, i.e. consider also the other effects predicted by this theory, like the precession of the nodes, the variation of the eccentricity and the inclination, and so on --- see Eqs.~\eqref{avperteqsI0}. Then one should perform a global fit of all these effects to the data; it is likely that in this way the quantitative conclusions would be different.

Finally let us cautiously remark that MOND and more sophisticated theories such as TeVeS~\cite{Bek04}, which are intended to describe the weak field regime of gravity (below $a_0$), may not be extrapolated without modification to the strong field of the Solar System. For instance it has been argued~\cite{FB05} that a MOND interpolating function $\mu$ which performs well at fitting the rotation curves of galaxies is given by $\mu_1$ defined by \eqref{mun}. However this function has a rather slow transition to the Newtonian regime, given by $\mu_1\sim 1-y^{-1}$ when $y=g/a_0\to\infty$, which is already excluded by Solar System observations. Indeed such slow fall-off $-y^{-1}$ predicts a constant supplementary acceleration directed toward the Sun $\delta g_\text{N}=a_0$ (i.e. a ``Pioneer'' effect), which is ruled out because not seen from the motion of planets. Thus it could be that the transition between MOND and the Newtonian regime is more complicated than what is modelled by Eq.~\eqref{e:MOND}. This is also true for the dipolar dark matter model~\cite{BL08,BL09} which may only give an effective description valid in the weak field limit and cannot be extrapolated as it stands to the Solar System. While looking at MOND-like effects in the Solar System we should keep the previous \textit{proviso} in mind. The potential conflict we find here with the Solar System dynamics (notably with the constraints on the orbital precession of Saturn~\cite{fienga09,fienga10}) may not necessarily invalidate those theories if they are not ``fundamental'' theories but rather ``phenomenological'' models only pertinent in a certain regime. 

In any case, further studies are to be done if one wants to obtain more stringent conclusions about constraints imposed by Solar-system observations onto MOND-like theories. More precise observations could give valuable informations about an eventual EFE due to the MOND theory and restrict the number of possible MOND functions that are compatible with the observations. More generally the influence of the Galactic field on the Solar-system dynamics through a possible violation of the strong version of the equivalence principle (of which the EFE is a by-product in the case of MOND) is worth to be investigated.

\bibliography{/home/blanchet/Articles/ListeRef/ListeRef.bib}
\end{document}